\documentclass[11pt,twoside]{article}

\usepackage{asp2006}
\usepackage{epsf}
\usepackage{psfig}
\usepackage{graphicx}
\usepackage{lscape}
\usepackage{amsbsy}
\usepackage{amssymb}

\begin{document}
\title{Photometric properties of magnetic elements: resolved and unresolved features}

\author{S. Criscuoli }   

\affil{INAF - Osservatorio astronomico di Roma , Via Frascati 33, I-00040, Monte Porzio Catone, Italia}

\begin {abstract} We investigate, by numerical simulations, the photometric signature of magnetic flux tubes in the solar photosphere. We show that the observed contrast profiles are determined not only by the physical properties  of the tube and its surroundings, but also by the peculiarities of the observations, including the line/continuum formation height and the spatial and spectral resolution. The aim is to understand these contributions well enough so that multi-wavelength observations can begin to disentangle them.

\end {abstract}

\section{Introduction}   
The high spatial resolution achieved by modern telescopes, along with improvements in real time and post-processing correction of image degradation, has allowed the detailed study of  the photometric properties of small magnetic bright points. Nevertheless, significant disparities exist in measurements of the contrast as a function of disk position (center-to-limb variation or CLV). Moreover, recent observations off disk center have shown that small magnetic features are often associated with the presence of 'dark lanes' and 'bright tails' \citep{lites2004, hirzberger2005, berger2007}, which are manifestations of thermal effects induced by the reduction of opacity inside the magnetic elements \citep{steiner2005}. In this paper, we demonstrate how the photometric signatures of small magnetic elements reflect, not only the physical properties of the flux tube and its surroundings, but also the peculiarities of the observation itself (such as spatial resolution and spectral range).

\begin{figure*}
\centering{
 \includegraphics[width=7.0cm, height=4.0cm ]{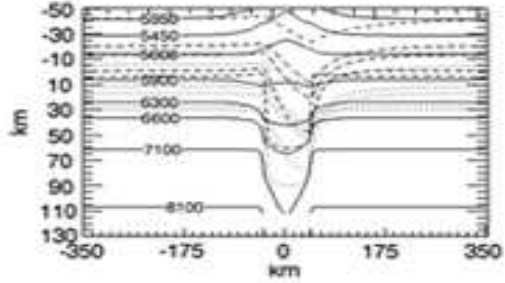}
\caption{Detail of the temperature field (in Kelvin) inside and around a magnetic flux tube. Tube diameter is 70km. The dotted lines represent the heights at which $\tau$=1 in the opacity model from Kurucz. The dashed lines represent the heights at which $\tau$=1  in the case of Kurucz opacity increased by a factor of two. For both cases iso-optical depths for  various lines of sight are shown: $\mu$=0.4,0.6,0.7,0.8,1 from top to bottom.}

\label{fig1}
}
\end{figure*}

\section{The Model}
We investigate, by 2D numerical simulations, the radiative properties of isolated and clustered static magnetic flux tubes in Radiative Equilibrium (RE, hereafter) with a surrounding plane parallel gray atmosphere. LTE is assumed. RE is imposed by an iterative scheme similar to the one proposed by \citet{fabiani1992}. The radiation intensity field is evaluated by a code (Criscuoli, 2007)\footnote{See http://dspace.uniroma2.it/dspace/advanced-search/} based on the short characteristic technique \citep{kunasz1988}. As initial conditions for atmospheric parameters, the Kurucz model \citep{kurucz1994} is adopted.
The presence of the magnetic field is mimicked by shifting downward the atmosphere at the flux tube location. The spatial domain of the simulations is 1.1$\times$2.3Mm with horizontal and vertical resolution of 3.5km and 7.1km, respectively. The magnetic field intensity is estimated, at each height, supposing the thin flux tube approximation \citep{stix2002}. 
An example of the simulated temperature field inside and around a flux tube in RE is given in fig.\ref{fig1}. Because of the reduction of opacity in the magnetic element, at large optical depths ($\tau \geq 1$) the temperature inside and immediately outside the tube is lower than the surroundings. At the shallower optical depths ($\tau < $  1), due to the radiative channeling effect \citep{fabiani1992}, the tube and the adjacent atmosphere are hotter.

\section{Formation height}

\begin{figure}
\centering{
 \includegraphics[width=5.0cm,height=3.5cm]{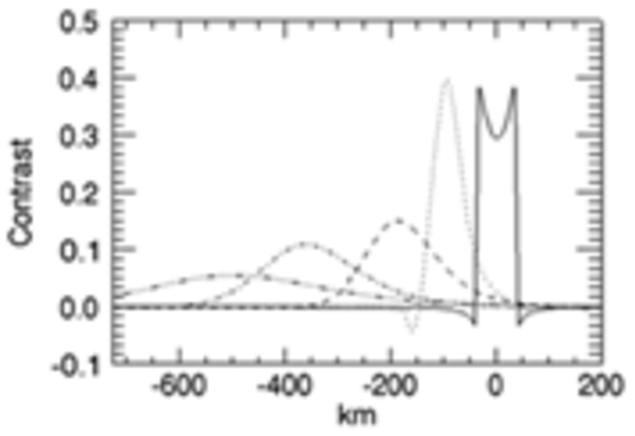}
\includegraphics[width=5.0cm,height=3.5cm]{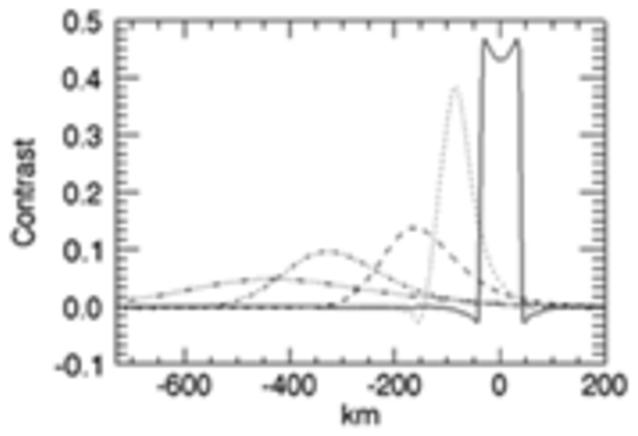}
\caption{Detail of contrast across the tube in fig.\ref{fig1} observed at two optical depths and different positions on the solar disk. Left: Kurucz opacity model. Right: Kurucz opacity increased by a factor of two. Continuous line: $\mu$=1. Dotted line: $\mu$=0.9. Dashed line: $\mu$=0.8. Dot-dashed line: $\mu$=0.7. Dot-dot-dashed line: $\mu$=0.4. Disk center is on the left. Curves have been shifted in order to fit into the same plot. 
}

\label{fig2}
}
\end{figure}

Because the thermal stratification depends on the height (see fig.\ref{fig1}), the contrast across magnetic elements is different when observed at wavelengths which sample different layers of the atmosphere. In our gray models this effect is investigated by comparing the contrast across a magnetic flux tube obtained by modifying the model adopted for the opacity. Figure \ref{fig1} shows $\tau$=1 lines at various lines of sight ($\mu$ is the cosine of the heliocentric angle) in the case of the Rosseland mean opacity from Kurucz model (dashed lines) and in the case in which the opacity is increased by a factor of two (dotted lines). Figure \ref{fig2} shows the corresponding  contrast profiles. At $\mu$=1 the Wilson depressions are approximately 60km in both cases, but the contrast is higher in the case of higher opacity, because the corresponding iso-optical line samples layers at which the temperature gradient is larger. Both plots show 'double humps' and 'dark rings', whose origin is discussed by \citet{knolker1988}. Because of the heating in the highest layers of the domain, the temperature difference between the 'wall' and the non magnetic atmosphere decreases monotonically with increasing the line of sight angle. Therefore, the maximum contrast value decreases monotonically with decreasing $\mu$ for the higher opacity model.


\begin{figure}
 \includegraphics[width=4.2cm,height=3.5cm]{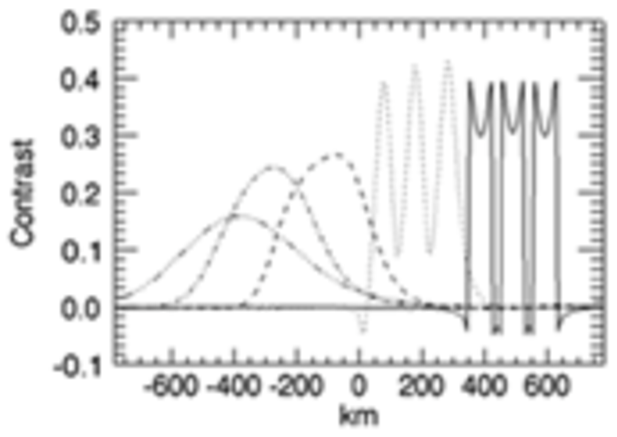}
\includegraphics[width=4.2cm,height=3.5cm]{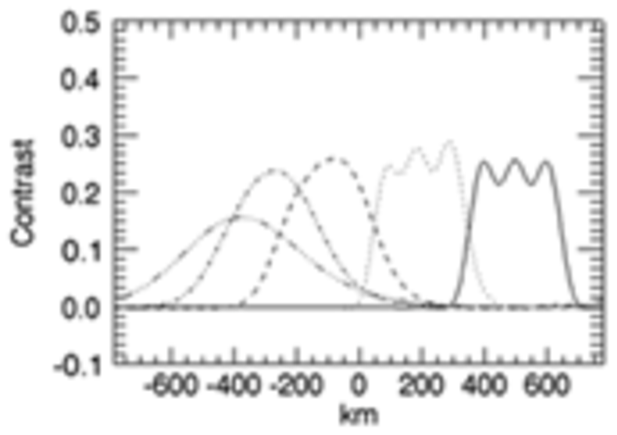}
\includegraphics[width=4.2cm,height=3.5cm]{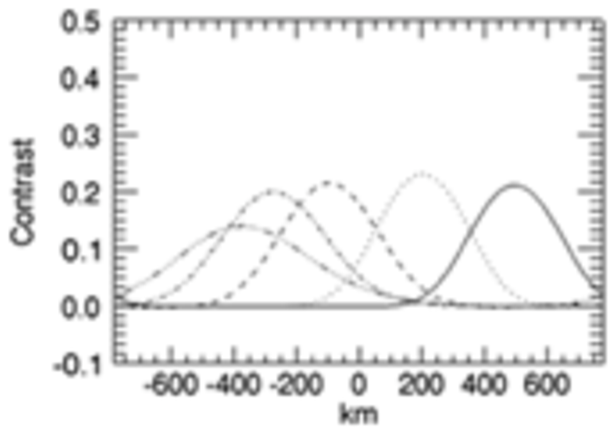}
\caption{Contrast profiles of three flux tubes (details in the text) observed at various positions on the solar disk and spatial resolutions. Left: Resolution of the simulation (3.5km). Center: 0.1". Right: 0.3". Legend is as in fig.\ref{fig2}.}
\label{fig3}
\end{figure}

\section{Resolution, Clustering and Identification methods}
Figure \ref{fig3} shows the contrast profiles observed at various positions on the solar disk and different resolutions for three flux tubes in RE with the surrounding atmosphere and whose mutual distances (from tube axis) and diameters are 105km and 70km, respectively. Boundary and initial conditions are as for the tube in fig.\ref{fig1}. The central and right plots show contrast profiles obtained convolving original profiles with Gaussian functions of FWHM=0.1" and FWHM=0.3", respectively. Figure \ref{fig3} shows that there is a 'critical line of sight' beyond which the flux tubes are not distinguishable and appear as a single structure. Comparison of fig.\ref{fig1} and \ref{fig3} shows that the contrast of clusters of magnetic elements is higher than the contrast of isolated elements. Finally, we notice that reduction of resolution mostly affects profiles observed at disk center, since these are characterized by features of spatial scales of tens of kilometers. 

\begin{figure*}
\centering{
 \includegraphics[width=5.0cm]{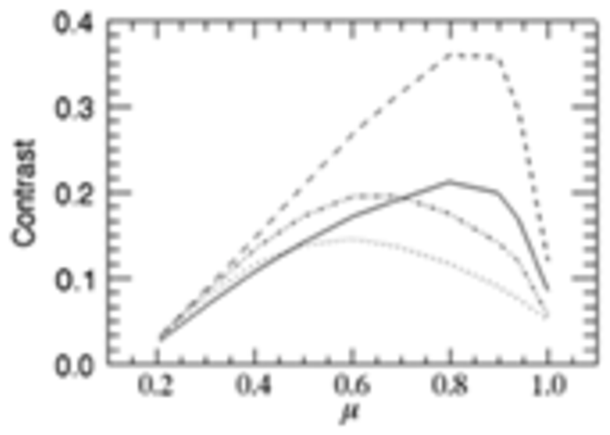}
\includegraphics[width=5.0cm]{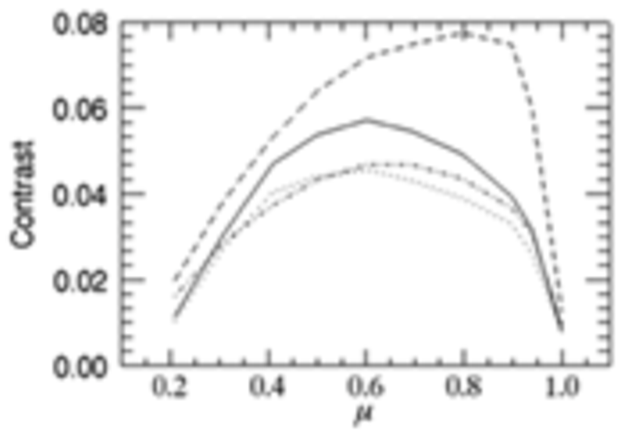}
\caption{CLV of contrasts of isolated (left) or three flux tubes (right) measured at different spatial resolution and with different methods. Boundary conditions are such that, when in RE, the magnetic flux measured with spatial resolution worse than 0.4" is approximately the same for the two configurations. Legend is explained in the text.}

\label{fig4}
}
\end{figure*}

In fig.\ref{fig4} we show CLV obtained by 4 different observing conditions and features definitions for an isolated flux tube of 175km diameter (left) and three flux tubes of 70km diameter and 105km apart (right):
1)Maximum contrast observed with spatial resolution 0.1" (dashed lines);
2)Mean contrast of elements whose contrast is larger than a threshold and observed with spatial resolution 0.1" (continuous line);
3)Mean contrast observed with spatial resolution 0.4";
4)Mean contrast observed with spatial resolution 0.6" (dotted line). 


The figure shows that the CLV of measured contrast of magnetic features depends on the image spatial resolution and on the technique employed to detect and define the features themselves. 
For instance, the two dotted curves of the two plots display different shapes and contrast values, although the magnetic field intensities of the three flux tubes are such that for resolution worse than 0.4" the magnetic flux measured is the same as for the isolated flux tube case. Therefore, methods based on magnetograms intensity thresholding would not distinguish between the single and the three flux tubes cases, thus producing a large spread of measured contrast.  

\section{Conclusions}
We have showed, by 2D simulations, that measured  intensity contrast of  magnetic elements depends not only on their physical properties, but also on the peculiarities  of the observations and on the data analysis techniques employed. Results obtained by this work will be useful for the interpretation of present and future high resolution observations, as well as for the interpretation of the discrepancies of some results presented in the literature.


\end{document}